\pdfoutput=1
\documentclass[12pt,a4paper]{article}

\usepackage{ifthen} 
\newboolean{pdflatex}
\setboolean{pdflatex}{true} 

\newboolean{articletitles}
\setboolean{articletitles}{true} 

\newboolean{uprightparticles}
\setboolean{uprightparticles}{false} 

\def\paperauthors{LHCb collaboration} 
\def\paperasciititle{Measurement of the nuclear modification factor and prompt charged particle production in pPb and pp collisions at 5 TeV} 
\def\papertitle{ Measurement of the \\nuclear modification factor and \\prompt charged particle production\\ in \pA and \pp collisions\\ at $\sqsnn=5\tev$ } 

\def\paperkeywords{{High Energy Physics}, {LHCb}} 
\def\papercopyright{\the\year\ CERN for the benefit of the LHCb collaboration} 
\def\paperlicence{CC BY 4.0 licence}
\def\paperlicenceurl{https://creativecommons.org/licenses/by/4.0/}

\usepackage[top=1in, bottom=1.25in, left=1in, right=1in]{geometry}

\usepackage{microtype}
\usepackage{lineno}  
\usepackage{xspace} 
\usepackage{caption} 

\usepackage{graphicx}  
\usepackage{color}
\usepackage{colortbl}
\graphicspath{{./figs/}} 

\usepackage{amsmath} 
\usepackage{amssymb}
\usepackage{amsfonts}
\usepackage{upgreek} 

\newcommand*\patchAmsMathEnvironmentForLineno[1]{%
\expandafter\let\csname old#1\expandafter\endcsname\csname #1\endcsname
\expandafter\let\csname oldend#1\expandafter\endcsname\csname
end#1\endcsname
 \renewenvironment{#1}%
   {\linenomath\csname old#1\endcsname}%
   {\csname oldend#1\endcsname\endlinenomath}%
}
\newcommand*\patchBothAmsMathEnvironmentsForLineno[1]{%
  \patchAmsMathEnvironmentForLineno{#1}%
  \patchAmsMathEnvironmentForLineno{#1*}%
}
\AtBeginDocument{%
\patchBothAmsMathEnvironmentsForLineno{equation}%
\patchBothAmsMathEnvironmentsForLineno{align}%
\patchBothAmsMathEnvironmentsForLineno{flalign}%
\patchBothAmsMathEnvironmentsForLineno{alignat}%
\patchBothAmsMathEnvironmentsForLineno{gather}%
\patchBothAmsMathEnvironmentsForLineno{multline}%
\patchBothAmsMathEnvironmentsForLineno{eqnarray}%
}

\usepackage{hyperxmp}

\usepackage[colorinlistoftodos,textsize=scriptsize]{todonotes}

\usepackage[bottom,flushmargin,hang,multiple]{footmisc}

\usepackage[pdftex,
            pdfauthor={\paperauthors},
            pdftitle={\paperasciititle},
            pdfkeywords={\paperkeywords},
            pdfcopyright={Copyright (C) \papercopyright},
            pdflicenseurl={\paperlicenceurl}]{hyperref}

\usepackage[all]{hypcap} 

\usepackage{xspace} 
\usepackage{upgreek}

\def\MagUp {\mbox{\em Mag\kern -0.05em Up}\xspace}

\ifthenelse{\boolean{uprightparticles}}%
{

 \def\PDelta      {\ensuremath{\Delta}\xspace}                 
 \def\PXi         {\ensuremath{\Xi}\xspace}                 
 \def\PLambda     {\ensuremath{\Lambda}\xspace}                 
 \def\PSigma      {\ensuremath{\Sigma}\xspace}                 
 \def\POmega      {\ensuremath{\Omega}\xspace}                 
 \def\PUpsilon    {\ensuremath{\Upsilon}\xspace}

 \def\PB      {\ensuremath{\mathrm{B}}\xspace}                 
                  
 \def\PD      {\ensuremath{\mathrm{D}}\xspace}

 \def\PK      {\ensuremath{\mathrm{K}}\xspace}

 \def\Pi      {\ensuremath{\mathrm{i}}\xspace}

 \def\Ps      {\ensuremath{\mathrm{s}}\xspace}

 \def\thebaroffset{0.0em}
}
{

 \mathchardef\PDelta="7101
 \mathchardef\PXi="7104
 \mathchardef\PLambda="7103
 \mathchardef\PSigma="7106
 \mathchardef\POmega="710A
 \mathchardef\PUpsilon="7107
                  
 \def\PB      {\ensuremath{B}\xspace}                 
                  
 \def\PD      {\ensuremath{D}\xspace}

 \def\PK      {\ensuremath{K}\xspace}

 \def\Pi      {\ensuremath{i}\xspace}

 \def\Ps      {\ensuremath{s}\xspace}

 \def\thebaroffset{0.18em}
}
\newcommand{\offsetoverline}[2][\thebaroffset]{\kern #1\overline{\kern -#1 #2}}%

\makeatletter
\ifcase \@ptsize \relax
  \newcommand{\miniscule}{\@setfontsize\miniscule{4}{5}}
\or
  \newcommand{\miniscule}{\@setfontsize\miniscule{5}{6}}
\or
  \newcommand{\miniscule}{\@setfontsize\miniscule{5}{6}}
\fi
\makeatother

\DeclareRobustCommand{\optbar}[1]{\shortstack{{\miniscule (\rule[.5ex]{1.25em}{.18mm})}
  \\ [-.7ex] $#1$}}

\def\squark    {{\ensuremath{\Ps}}\xspace}

\def\kaon    {{\ensuremath{\PK}}\xspace}

\def\KorKbar {\kern \thebaroffset\optbar{\kern -\thebaroffset \PK}{}\xspace}

\def\Kp      {{\ensuremath{\kaon^+}}\xspace}
\def\Km      {{\ensuremath{\kaon^-}}\xspace}

\def\KS      {{\ensuremath{\kaon^0_{\mathrm{S}}}}\xspace}

\def\D       {{\ensuremath{\PD}}\xspace}

\def\DorDbar {\kern \thebaroffset\optbar{\kern -\thebaroffset \PD}\xspace}

\def\Dp      {{\ensuremath{\D^+}}\xspace}
\def\Dm      {{\ensuremath{\D^-}}\xspace}

\def\DpDm    {\ensuremath{\Dp {\kern -0.16em \Dm}}\xspace}

\def\B       {{\ensuremath{\PB}}\xspace}

\def\BorBbar {\kern \thebaroffset\optbar{\kern -\thebaroffset \PB}\xspace}

\def\Bd      {{\ensuremath{\B^0}}\xspace}

\def\BdorBdbar {\kern \thebaroffset\optbar{\kern -\thebaroffset \Bd}\xspace}

\def\Bs      {{\ensuremath{\B^0_\squark}}\xspace}

\def\BsorBsbar {\kern \thebaroffset\optbar{\kern -\thebaroffset \Bs}\xspace}

\def\Y#1S{\ensuremath{\PUpsilon{(#1S)}}\xspace}

\def\Lz          {{\ensuremath{\PLambda}}\xspace}

\def\LorLbar     {\kern \thebaroffset\optbar{\kern -\thebaroffset \PLambda}\xspace}

\def\to                 {\ensuremath{\rightarrow}\xspace}

\def\AT#1     {\ensuremath{A_{\mathrm{T}}^{#1}}\xspace}           

\def\C#1      {\ensuremath{\mathcal{C}_{#1}}\xspace}                       
\def\Cp#1     {\ensuremath{\mathcal{C}_{#1}^{'}}\xspace}                    
\def\Ceff#1   {\ensuremath{\mathcal{C}_{#1}^{\mathrm{(eff)}}}\xspace}        
\def\Cpeff#1  {\ensuremath{\mathcal{C}_{#1}^{'\mathrm{(eff)}}}\xspace}       
\def\Ope#1    {\ensuremath{\mathcal{O}_{#1}}\xspace}                       
\def\Opep#1   {\ensuremath{\mathcal{O}_{#1}^{'}}\xspace}                    

\newcommand{\nospaceunit}[1]{\ensuremath{\text{#1}}}       
\newcommand{\aunit}[1]{\ensuremath{\text{\,#1}}}       

\newcommand{\tev}{\aunit{Te\kern -0.1em V}\xspace}
\newcommand{\gev}{\aunit{Ge\kern -0.1em V}\xspace}
\newcommand{\mev}{\aunit{Me\kern -0.1em V}\xspace}
\newcommand{\kev}{\aunit{ke\kern -0.1em V}\xspace}
\newcommand{\ev}{\aunit{e\kern -0.1em V}\xspace}
 
\newcommand{\mevc}{\ensuremath{\aunit{Me\kern -0.1em V\!/}c}\xspace}
\newcommand{\gevc}{\ensuremath{\aunit{Ge\kern -0.1em V\!/}c}\xspace}
\newcommand{\mevcc}{\ensuremath{\aunit{Me\kern -0.1em V\!/}c^2}\xspace}
\newcommand{\gevcc}{\ensuremath{\aunit{Ge\kern -0.1em V\!/}c^2}\xspace}

\def\mbarn{\aunit{mb}\xspace}
\def\mub{\ensuremath{\,\upmu\nospaceunit{b}}\xspace}
\def\nb {\aunit{nb}\xspace}
\def\invnb {\ensuremath{\nb^{-1}}\xspace}
\def\pb {\aunit{pb}\xspace}
\def\invpb {\ensuremath{\pb^{-1}}\xspace}

\def\sec  {\ensuremath{\aunit{s}}\xspace}

\newcommand{\chisq}{\ensuremath{\chi^2}\xspace}

\def\deriv {\ensuremath{\mathrm{d}}}

\def\gsim{{~\raise.15em\hbox{$>$}\kern-.85em
          \lower.35em\hbox{$\sim$}~}\xspace}
\def\lsim{{~\raise.15em\hbox{$<$}\kern-.85em
          \lower.35em\hbox{$\sim$}~}\xspace}

\def\sqsnn {\ensuremath{\protect\sqrt{s_{\scriptscriptstyle\text{NN}}}}\xspace}
\def\pt         {\ensuremath{p_{\mathrm{T}}}\xspace}

\def\ptot       {\ensuremath{p}\xspace}

\newcommand{\lum} {\ensuremath{\mathcal{L}}\xspace}

\def\evtgen     {\mbox{\textsc{EvtGen}}\xspace}

\def\geant      {\mbox{\textsc{Geant4}}\xspace}

\def\pythia     {\mbox{\textsc{Pythia}}\xspace}

\def\tell1  {TELL1\xspace}
\def\ukl1   {UKL1\xspace}

\usepackage{cite} 
\usepackage{mciteplus}

\input{local-symbols-def} 
\usepackage{longtable}
\usepackage{arydshln}
\usepackage{multirow}

\begin{document}

\renewcommand{\thefootnote}{\fnsymbol{footnote}}
\setcounter{footnote}{1}


\begin{titlepage}
\pagenumbering{roman}

\vspace*{-1.5cm}
\centerline{\large EUROPEAN ORGANIZATION FOR NUCLEAR RESEARCH (CERN)}
\vspace*{1.5cm}
\noindent
\begin{tabular*}{\linewidth}{lc@{\extracolsep{\fill}}r@{\extracolsep{0pt}}}
\ifthenelse{\boolean{pdflatex}}
{\vspace*{-1.5cm}\mbox{\!\!\!\includegraphics[width=.14\textwidth]{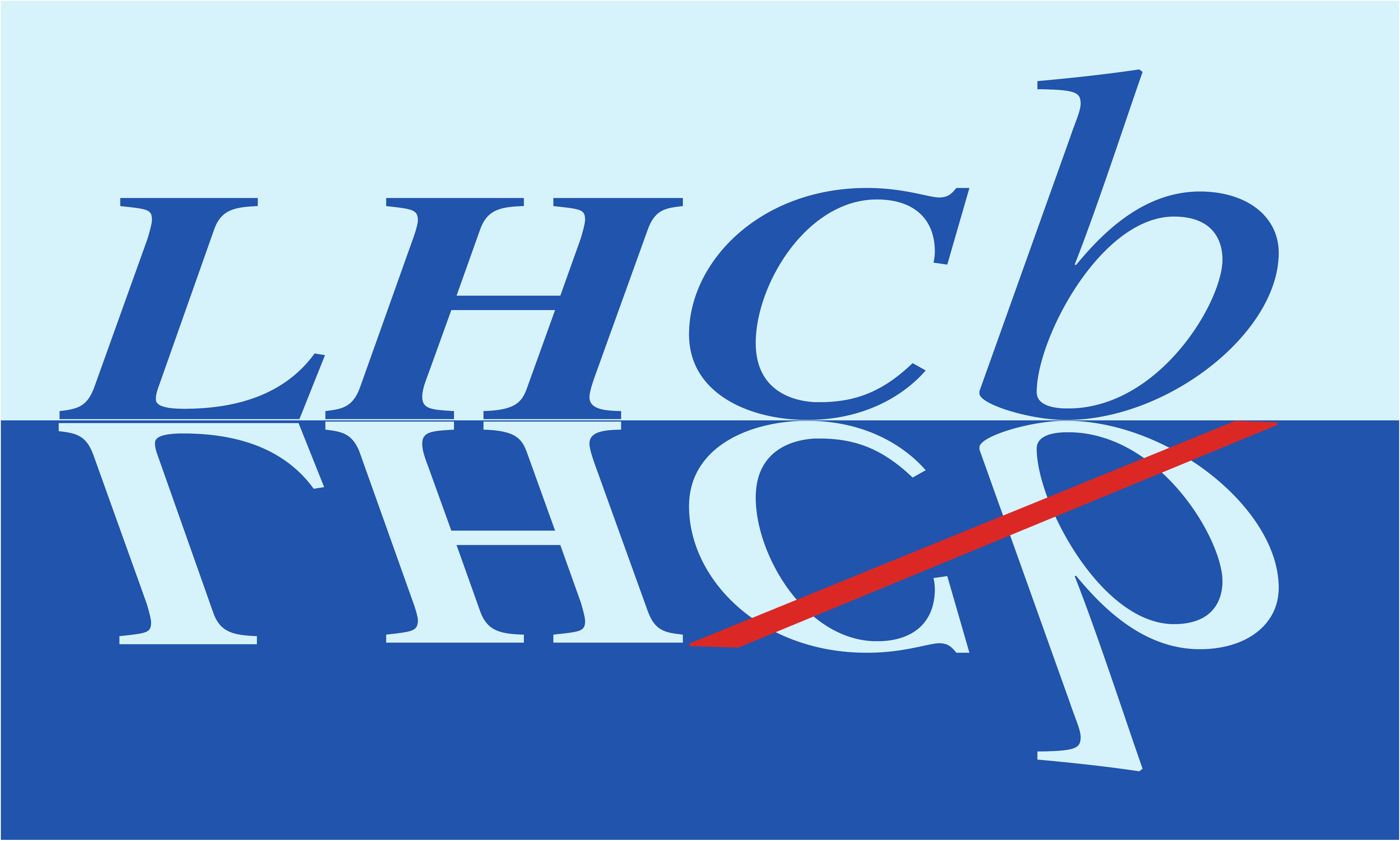}} & &}%
{\vspace*{-1.2cm}\mbox{\!\!\!\includegraphics[width=.12\textwidth]{figs/lhcb-logo.eps}} & &}%
\\
 & & CERN-EP-2021-130 \\  
 & & LHCb-PAPER-2021-015 \\  
 & & June 21, 2022 \\ 
 & & \\
\end{tabular*}

\vspace*{3.0cm}

{\normalfont\bfseries\boldmath\huge
\begin{center}
  \papertitle 
\end{center}
}

\vspace*{1cm}

\begin{center}
\paperauthors\footnote{Authors are listed at the end of this paper.}
\end{center}

\vspace{\fill}

\begin{abstract}
  \noindent
The production of prompt charged particles in proton-lead collisions and in proton-proton collisions at the nucleon-nucleon centre-of-mass energy ${\sqsnn=5\tev}$ is studied at LHCb as a function of pseudorapidity ($\eta$) and transverse momentum ($\pt$) with respect to the proton beam direction. The nuclear modification factor for charged particles is determined as a function of $\eta$ between ${-4.8<\eta<-2.5}$ (backward region) and ${2.0<\eta<4.8}$ (forward region), and $\pt$ between ${0.2<\pt<8.0\gevc}$. The results show a suppression of charged particle production in proton-lead collisions relative to
proton-proton collisions in the forward region and an enhancement in the backward region for $\pt$ larger than $1.5\gevc$. This measurement constrains nuclear PDFs and saturation models at previously unexplored values of the parton momentum fraction down to $10^{-6}$.
\end{abstract}

\vspace*{0.5cm}

\begin{center}
    Published in Phys.Rev.Lett. 128 (2022), 142004 
\end{center}

\vspace{\fill}

{\footnotesize 
\centerline{\copyright~\papercopyright. \href{\paperlicenceurl}{\paperlicence}.}}
\vspace*{2mm}

\end{titlepage}


\newpage
\setcounter{page}{2}
\mbox{~}
%
%
%
%


\renewcommand{\thefootnote}{\arabic{footnote}}
\setcounter{footnote}{0}

\cleardoublepage


\pagestyle{plain} 
\setcounter{page}{1}
\pagenumbering{arabic}


Charged particle production in hadronic collisions is a fundamental observable for studying the properties of the strong interaction governed by quantum chromodynamics (QCD). In high-energy collisions at the Large Hadron Collider (LHC), charged particles can be produced in soft and hard interactions which correspond to small and large momentum exchanges between the interacting partons of the hadrons, respectively. While hard interactions can be described by perturbative QCD (pQCD), the soft regime is less well understood and predictions currently rely on phenomenological considerations~\cite{Kaidalov:1983ew,CAPELLA1994225}. Even at LHC energies, charged particles from soft interactions dominate over those from hard interactions. For this reason, experimental input is crucial to improve models and generators for hadron collider and cosmic ray physics~\cite{Buckley:2011ms,dEnterria:2011twh,Albrecht:2021cxw}.

The study of the hard regime, which corresponds to charged particles of high transverse momentum (\pt) with respect to the axis of the colliding hadrons, provides valuable information on the physics of heavy-ion collisions~\cite{Salgado:2011wc}. Modifications of the charged particle production rate in proton-lead ($\pA$) collisions relative to proton-proton ($\pp$) collisions can be modelled assuming a variety of cold nuclear matter (CNM) effects~\cite{Albacete:2013ei,Albacete:2017qng}. Recent indications of collective fluid-like phenomena in small systems suggest the presence of dynamics not generally classified as CNM effects but as signatures of a quark gluon plasma~\cite{Nagle:2018nvi}.

For charged particles, these modifications are generally associated with initial-state effects, parameterised in nuclear parton distribution functions (nPDFs)~\cite{npdfEPPS16,nCTEQ15-Kovarik:2015cma,deFlorian:2011fp}. Other nuclear effects are related to initial- or final-state multiple scatterings of incoming and outgoing partons~\cite{Kang:2013ufa,Accardi:2002ik}, and could manifest in a Cronin enhancement~\cite{Cronin:1974zm}. Another approach considers models based on parton saturation, an effect arising at low values of the parton momentum fraction, $x$, and heavy nuclei~\cite{Kowalski:2007rw}. In this regime, the QCD dynamics can be described by the colour glass condensate (CGC) effective field theory~\cite{CGCMcLerran:1993ka}. Pion production at central rapidity~\cite{ALICE:2021est} is well described by modified nPDFs, energy loss and CGC calculations~\cite{npdfEPPS16,nCTEQ15-Kovarik:2015cma,Arleo:2020hat,Lappi:2013zma}. Low values of $x$, where saturation effects are most likely to occur, can be probed with high-energy collisions at the most forward rapidities.

Previous studies at the LHC~\cite{ALICEchpart2018,CMSchpart2017,ATLASchpart2016} have measured prompt charged particle production in \pA collisions at the centre-of-mass energy in the nucleon-nucleon system ${\sqsnn=5\tev}$ in the central pseudorapidity region. At the Relativistic Heavy Ion Collider (RHIC), measurements with deuteron-gold and proton-gold ($p\mathrm{Au}$) collisions at more forward rapidities but lower energy (${\sqsnn=200\gev}$) have been performed~\cite{BRAHMSchpart2004,PHENIXchpart2020,Back:2004bq}. The LHCb experiment can uniquely probe the lowest $x$ ranges currently accessible, given its forward rapidity coverage and higher collision energy.

This Letter presents the measurement of the prompt charged particle spectra in \pA and \pp collisions at ${\sqsnn=5\tev}$ in the ${0.2<\pt<8.0\gevc}$ range, thus covering the soft and hard production regimes. The \pA measurement covers the \textit{backward} pseudorapidity range of ${-5.2<\eta<-2.5}$, where the lead beam enters the LHCb spectrometer at the interaction point, and the \textit{forward} pseudorapidity range of ${1.6<\eta<4.3}$, where the proton beam enters the LHCb spectrometer at the interaction point. The $\pp$ measurement spans over ${2.0<\eta<4.8}$, and complements the recent measurement of prompt charged particle production at ${\sqsnn=13\tev}$~\cite{LHCb-PAPER-2021-010}. Throughout the text $\eta$ is expressed in the nucleon-nucleon centre-of-mass system, and is related with the pseudorapidity in the laboratory frame $\eta_{\mathrm{lab}}$ by ${\eta = \eta_{\mathrm{lab}} - 0.465}$ for \pA and ${\eta=\eta_{\mathrm{lab}}}$ for $\pp$ collisions.

The double-differential production cross-section for prompt charged particles is measured as
\begin{equation}
\cfrac{ \deriv^2\sigma^{\mathrm{ch}}(\eta,\pt) }{\deriv \pt \deriv\eta}\equiv\dfrac{1}{\lum } \dfrac{N^{\mathrm{ch}}(\eta, \pt)}{\Delta \pt \Delta \eta}
.
\label{eq:cross-section}    
\end{equation}
Here, $N^{\mathrm{ch}}$ is the number of prompt charged particles produced in a given interval of $\eta$ and $\pt$, $\Delta \eta$ and $\Delta \pt$, respectively, and $\lum$ is the integrated luminosity of the corresponding data sample. In this study a prompt charged particle is any charged hadron or lepton with a mean lifetime above ${0.3\times10^{-10}\sec}$ produced directly in the collision or from decays of shorter-lifetime particles~\cite{ALICEprimarypart}. The nuclear modification factor, \Rppb, is defined as
\begin{equation}
\Rppb(\eta,\pt)\equiv\cfrac{1}{A}\cfrac{ \deriv^2\sigma^{\mathrm{ch}}_{ p \mathrm{Pb}}(\eta,\pt)/\deriv \pt \deriv\eta }{ \deriv^2\sigma^{\mathrm{ch}}_{pp}(\eta,\pt)/\deriv \pt \deriv \eta}
,
\label{eq:RpPb}
\end{equation}
where ${A=208}$ is the number of nucleons in the lead ion and ${\deriv^2\sigma^{\mathrm{ch}}_{\pA,\pp}(\eta,\pt)/\deriv \pt \deriv \eta}$ is the double-differential cross-section in $\pA$ and $\pp$ collisions, respectively.

The LHCb detector is a single-arm forward spectrometer described in Refs.~\cite{LHCb-DP-2008-001,LHCb-DP-2014-002}. The detector elements that are particularly relevant to this analysis are a silicon-strip vertex detector (VELO) surrounding the interaction region that allows the determination of the position of the collision point, known as the primary vertex (PV), a tracking system that provides a measurement of the momentum, \ptot, of charged particles and two ring-imaging Cherenkov detectors that are able to discriminate between different species of charged particles.

The corresponding integrated luminosity for the forward (backward) $\pA$ data sample is ${42.7\pm1.0\mub^{-1}}$ (${38.7\pm1.0\mub^{-1}}$)~\cite{LHCb-PAPER-2014-047}, where the uncertainties are uncorrelated between the two data samples. Events are required to pass a minimum-bias trigger which requires at least one reconstructed track in the VELO detector. Additionally, only events with one reconstructed PV within three standard deviations from the mean PV position of the full sample are considered.

The \pp data correspond to an integrated luminosity of ${3.49 \pm 0.07\invnb}$. An unbiased trigger for \pp events is used to select every leading bunch crossing that occurred during the data taking period, thus avoiding potential contamination between neighbouring bunches.

Simulation is used to model the reconstruction efficiency, the effects of the selection requirements, and the contribution from background tracks. In the simulation, $\pA$ collisions are generated using EPOS-LHC~\cite{EPOSLHC}, while $\pp$ collisions are generated using \pythia~\cite{Sjostrand:2007gs} with a specific LHCb configuration~\cite{LHCb-PROC-2010-056}. Particle decays are described by \evtgen~\cite{Lange:2001uf}, while the interaction of particles with the detector, and its response in simulation, are implemented using the \geant toolkit~\cite{Agostinelli:2002hh,Allison:2006ve} as described in Ref.~\cite{LHCb-PROC-2011-006}.

Prompt charged particle candidates are defined as tracks with hits in the VELO and the tracking stations after the LHCb detector dipole. This last condition requires the measured particle to have $\ptot>2\gevc$. Background contributions due to fake tracks and secondary particles are considered in this study. Fake tracks are reconstruction artefacts that do not correspond to actual charged particles, which are particularly relevant for large detector occupancies and in the high-$\pt$ region. Secondary particles are tracks produced by charged particles that do not meet the prompt particle definition and originate from interactions of particles with the detector material or from decays of prompt particles.

A selection is applied to reduce these background contributions. Fake tracks are suppressed with a tight requirement on the output of a neural-net based algorithm (ghost probability)~\cite{DeCian:2255039}. To suppress further the fake track background, when two or more candidates from the same event share a segment reconstructed in the VELO, only the candidate with the best track fit-quality is retained. Secondary particles are reduced by requiring small impact parameters with respect to the mean of the PV distribution in the full sample. This criterion is particularly effective at removing hadrons produced in decays of prompt $\KS$ and $\Lz$ particles and in interactions of hadrons with the detector material without inducing a bias by requiring a PV. 

The prompt charged particle yield, $N^{\mathrm{ch}}$, for the $\pA$ and $\pp$ samples is obtained from the number of candidates, which is corrected with the reconstruction efficiency, the selection efficiency and the signal purity. The reconstruction efficiency accounts for detector inefficiencies or acceptance effects. The track-finding efficiency from simulation is corrected with a tag-and-probe method applied to data and simulation in two-dimensional intervals of $\eta_{\mathrm{lab}}$ and \pt using \mbox{$J/\psi\rightarrow \mu^+\mu^-$} decays in the range ${5<\ptot<200\gevc}$~\cite{LHCb-DP-2013-002}. Since the reconstruction efficiency depends on the detector occupancy, the simulated samples are weighted to reproduce the occupancy distributions from different LHCb subdetectors in data. Additionally, the reconstruction efficiency depends on the particle type. The relative abundance of particles determined from simulation are validated with data from the ALICE~\cite{ALICEpionKprotonpPb5TeV2016,ALICEstrangenesspp7TeV,ALICExiomegapPb5TeV2016,ALICEpionKprotonLambdaKspPb5TeV2013} and LHCb~\cite{LHCb-PAPER-2011-037} experiments. The LHCb \pythia tune for $\pp$ collisions does not reproduce the kaon and prompt hyperons relative abundance at high $\pt$ in data. Therefore, a dedicated simulated sample generated with EPOS-LHC~\cite{EPOSLHC} is used to parameterise the particle composition in $\pp$ collisions. The relative abundances produced with the EPOS-LHC generator are in agreement with the data within $30\%$ in $\pp$ and $\pA$ collisions. As a cross-check, the relative abundances from EPOS-LHC in the LHCb acceptance are found to be compatible with those produced with \pythia using the rope hadronisation model~\cite{Sjostrand:2014zea,PYTHIARopesPaper}. The use of the uncorrected LHCb \pythia tune would imply an overestimation of the reconstruction efficiency up to a $7\%$ at high $\pt$.

The selection efficiency accounts for the fraction of prompt charged particles removed from the candidate sample by the selection. The efficiency is estimated using simulation and a dedicated calibration procedure using a tag-and-probe technique applied to ${\phi(1020)\to\Kp\Km}$ decays in data and simulation.

The signal purity is determined in simulation and corrected with background-enriched samples of data. Two independent samples dominated by fake tracks are constructed: using tracks with high ghost probability, and tracks which share their reconstructed VELO segment with a better fit-quality track. For secondary particles, the  contributions from hadrons and electrons are studied separately. An enriched sample of hadrons from prompt-particle decays, such as $\Lz$ baryons and $\KS$ mesons, and hadrons produced in material interactions is obtained using tracks with a large estimated effective impact parameter with respect to the mean of the PV distribution. The abundance of electrons from $\gamma$ conversions, which is considerable at low $\pt$, is validated using particle identification detectors.
 
Additionally, bin migration effects due to the
resolution of the detector are found to have a negligible contribution to the measured yields. A minor correction is made since the mass of the particle is ignored in the expression ${\eta = \eta_{\mathrm{lab}} - 0.465}$ which is used to translate the pseudorapidity in the laboratory system to the centre-of-mass system of the nucleon-nucleon collision in \pA collisions.

Several sources of systematic uncertainty are considered. For candidates in the range ${5<\ptot<200\gevc}$ the track-finding efficiency carries an uncertainty due to the limited size of the calibration samples and the difference between hadron and muon material interactions. For candidates outside this range, a conservative $5\%$ uncertainty is assigned based on the efficiency variation in adjacent intervals. An uncertainty is assigned accounting for the detector occupancy description, which is estimated considering alternative weights. The uncertainty due to imperfect knowledge of the relative particle composition is determined from a $30\%$ variation in the relative abundances of particles obtained from simulation. The uncertainty on the selection efficiency originates primarily from the limited size of the calibration sample. For the purity, the systematic uncertainty is estimated from the background abundance in the background-enriched samples and the data-simulation discrepancy in the background fraction from the independent samples. This uncertainty has a large $\eta$ and $\pt$ dependence: while negligible in regions with a small background level, it is the dominant contribution for intervals with large background contributions. These intervals correspond to high $\pt$ for fake tracks in $\pA$ collisions in the backward region. The uncertainties are given in \Tabref{tab:sysUncSummary}, where all uncertainties are treated as uncorrelated.

\begin{table}[t!]
\centering
\caption{ Relative uncertainties for \pA and \pp charged particle cross-sections. The interval indicates the minimum and the maximum value among the $(\eta, \pt)$ bins. The systematic uncertainty due to luminosity is fully correlated across the bins. The other sources of systematic uncertainties are fully uncorrelated between different bins.  }
\label{tab:sysUncSummary}
\begin{tabular}{lccc}
\hline \multirow{2}{*}{Uncertainty source}              & $\pA$[$\%$]                     & $\pA$[$\%$]                     & \multirow{2}{*}{\pp[$\%$]}                       
\\
                                & (forward)                     &  (backward)                      &                     \\\hline
Track-finding efficiency        & 1.5$\phantom{1}$--$\phantom{1}$5.0   & 1.5$\phantom{1}$--$\phantom{1}$5.0   & 1.6$\phantom{1}$--$\phantom{1}$5.3   \\
Detector occupancy              & 0.0$\phantom{1}$--$\phantom{1}$2.8   & 0.6$\phantom{1}$--$\phantom{1}$2.9   & 0.1$\phantom{1}$--$\phantom{1}$1.6   \\
Particle composition            & 0.4$\phantom{1}$--$\phantom{1}$4.1   & 0.4$\phantom{1}$--$\phantom{1}$4.6   & 0.3$\phantom{1}$--$\phantom{1}$2.4   \\
Selection efficiency            & 0.7$\phantom{1}$--$\phantom{1}$2.2   & 0.7$\phantom{1}$--$\phantom{1}$3.0   & 1.0$\phantom{1}$--$\phantom{1}$1.7   \\
Purity                          & 0.1$\phantom{1}$--$\phantom{1}$1.8   & 0.1$\phantom{1}$--11.7               & 0.1$\phantom{1}$--$\phantom{1}$5.8   \\
Truth-matching                  & 0.0$\phantom{1}$--$\phantom{1}$0.1   & 0.0$\phantom{1}$--$\phantom{1}$0.1   & 0.1$\phantom{1}$--$\phantom{1}$0.2   \\
Luminosity          & 2.3       & 2.5       & 2.0       \\
Statistical uncertainty         & 0.0$\phantom{1}$--$\phantom{1}$0.6   & 0.0$\phantom{1}$--$\phantom{1}$1.0   & 0.0$\phantom{1}$--$\phantom{1}$1.1   \\ \hline
 Total (in $\deriv^{2}\sigma/\deriv\eta \deriv\pt$) & 3.0$\phantom{1}$--$\phantom{1}$6.7   & 3.3$\phantom{1}$--14.5  & 2.8$\phantom{1}$--$\phantom{1}$8.7   \\
 \hline
Total (in $\Rppb$)                  & 4.2$\phantom{1}$--$\phantom{1}$9.2                     & 4.4$\phantom{1}$--16.9                    &                             \\ \hline
\end{tabular}
\end{table}

\begin{figure*}[!t]
\begin{center}
\includegraphics[width=1.\linewidth]{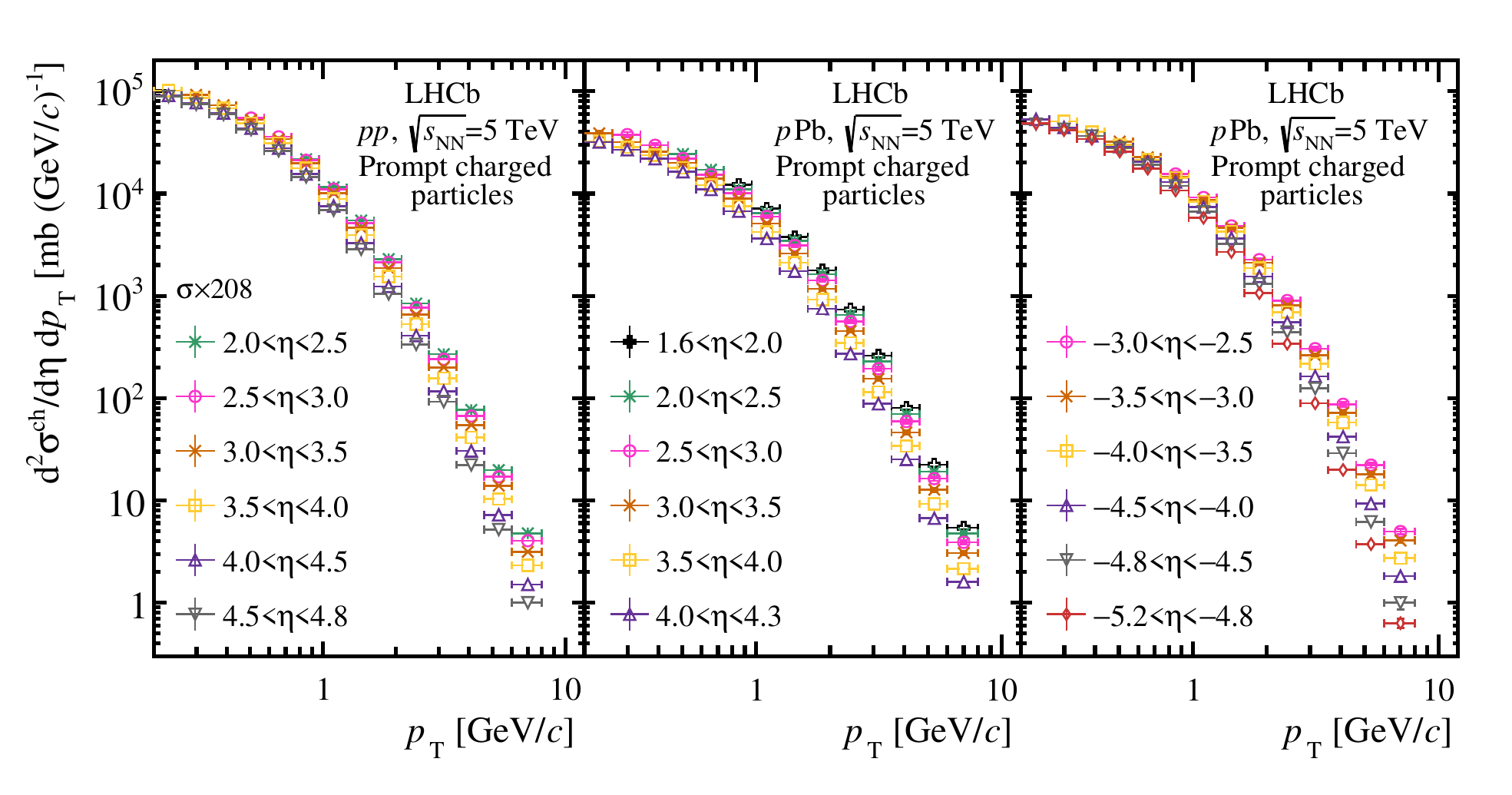}
\end{center}
\caption{ Differential cross-section of prompt charged particle production as a function of \pt in different $\eta$ intervals in (left) $\pp$, (middle) forward $\pA$ and (right) backward $\pA$ collisions. The cross-section values for $\pp$ are scaled by the lead mass number (${A=208}$) for comparison with the $\pA$ cross-sections. }
\label{fig:xSectionResults}
\end{figure*}

The measured prompt charged particle cross-sections for $\pp$ and $\pA$ are presented in \Figref{fig:xSectionResults}. The total uncertainty is the sum in quadrature of statistical, systematic, and luminosity uncertainties. On average ${0.1035\pm0.0029}$ charged particles (with ${0.961<\pt<1.249\gevc}$ and ${3.0<\eta<3.5}$) are produced in \pp collisions, when scaled by the total inelastic nucleon-nucleon cross-section of ${67.6\pm0.6\mbarn}$ at ${\sqsnn=5\tev}$~\cite{Loizides:2017ack}. This is two orders of magnitude smaller than for $\pA$ collisions, assuming the same total inelastic nucleon-nucleon cross-section. The cross-section for $\pp$ collisions at ${\sqsnn=5\tev}$ is compared with the result at ${\sqsnn=13\tev}$~\cite{LHCb-PAPER-2021-010}. Both results are consistent, showing an increase in the cross-section at $13\tev$ of a factor 1 to 3, depending on \pt.
 
\begin{figure*}[t]
\begin{center}
\includegraphics[width=1\linewidth]{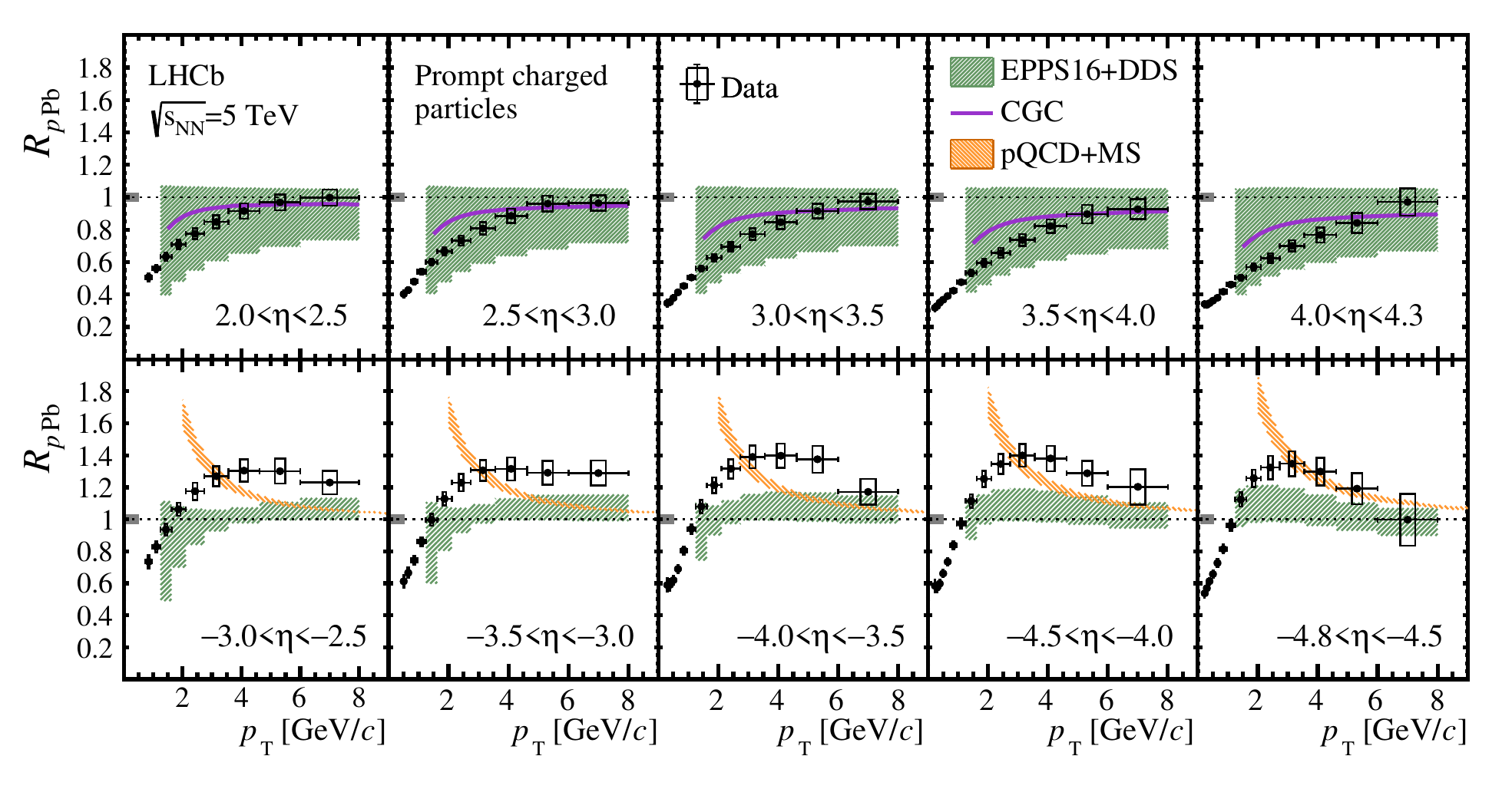}
\end{center}
\caption{ Nuclear modification factor as a function of $\pt$ in different $\eta$ intervals for the (top) forward and (bottom) backward regions, compared with the predictions from Refs.~\cite{Helenius:2014qla,Lappi:2013zma,Kang:2013ufa,Kang:2014hha}. Vertical error bars correspond to statistical uncertainties, open boxes to uncorrelated systematic uncertainty and the filled box at ${\Rppb=1}$ to the correlated uncertainty from the luminosity. }
\label{fig:RpA_RAptheory}
\end{figure*}

The result for $\Rppb$ in different ${(\eta,\pt})$ intervals is presented in \Figref{fig:RpA_RAptheory}, where the uncertainties arise from statistical, systematic and luminosity sources. In the forward region, the measurement indicates a suppression of charged particle production in \pA collisions relative to that in \pp collisions, which increases towards forward pseudorapidities. In the low $\pt$ regime, $\Rppb$ reaches values of about $0.3$ in the most forward pseudorapidities. In the backward region, a significant enhancement of $\Rppb$ is observed for ${\pt>1.5\gevc}$. This can be interpreted as Cronin enhancement~\cite{Cronin:1974zm}. The enhancement reaches a maximum at different $\pt$ values depending on $\eta$, followed by a decreasing trend towards unity. This decrease is more pronounced in the most backward pseudorapidities. The maximum value of \Rppb is found to be $\sim1.3$ and depends slightly on $\eta$.

The $\Rppb$ measurement is compared in \Figref{fig:RpA_RAptheory} with predictions from phenomenological models covering the ${\pt\gtrsim 1.5\gevc}$ region. The prediction in Ref.~\cite{Helenius:2014qla} (shaded green) is based on the nPDF set EPPS16~\cite{npdfEPPS16} for the lead nucleus and the PDF set CT14~\cite{Dulat:2015mca} for the proton. The calculation also employs the parton-to-hadron fragmentation functions set DSS~\cite{deFlorian:2007aj}. The prediction reproduces the data in the forward region although with large uncertainties. However, it fails to reproduce the $\Rppb$ enhancement in the backward region for ${\pt>2\gevc}$.

The second prediction (violet) is based on the CGC effective field theory~\cite{Lappi:2013zma}. The model is only applicable to the saturation region at low-$x$ and thus to forward rapidities. The predicted gradual decrease of $\Rppb$ with $\eta$ is observed in the data, although the prediction overestimates $\Rppb$ in the lower $\pt$ intervals. The prediction does not include an uncertainty estimation.

The third prediction (shaded orange) is a pQCD calculation within the high-twist factorisation formalism in the backward region~\cite{Kang:2013ufa,Kang:2014hha}. The calculation shows an enhancement due to incoherent multiple scattering inside the nucleus before and after the hard scattering, and reproduces the enhancement seen in $p\mathrm{Au}$ collisions in the backward region by the PHENIX experiment at ${\sqsnn=200\gev}$~\cite{PHENIXchpart2020}. The prediction shows a $\pt$ trend similar to data for ${\pt>3\gevc}$ in the most backward $\eta$ interval, although it does not reproduce the data for the other intervals in the backward configuration.

Understanding the evolution of $\Rppb$ with $x$ and the momentum transfer $Q^{2}$, is a critical point for the study of CNM effects. However, $x$ and $Q^{2}$ are partonic quantities and cannot be directly measured. Instead, experimental proxies for $x$ and $Q^2$~\cite{Armesto:2006ph}, defined as
\begin{equation}
Q^{2}_{\text{exp}}\equiv m^{2}+\pt^{2}\;\;\;\mathrm{and}\;\;\;x_{\text{exp}}\equiv \frac{Q_{\text{exp}}}{\sqsnn}e^{-\eta},
\label{eq:xexperiment}
\end{equation}
are considered to compare the \Rppb results among different LHC experiments. Here, $m$ is the mass of the produced particle and is taken as ${m=256\mevcc}$, the average charged particle mass in $\pA$ collisions determined with EPOS-LHC. The variable $x_{\text{exp}}$ is approximately $x$ for a two body scattering, and $Q_{\text{exp}}$ is the transverse mass of the produced particle.

\begin{figure}[t]
\begin{center}
\includegraphics[width=\linewidth]{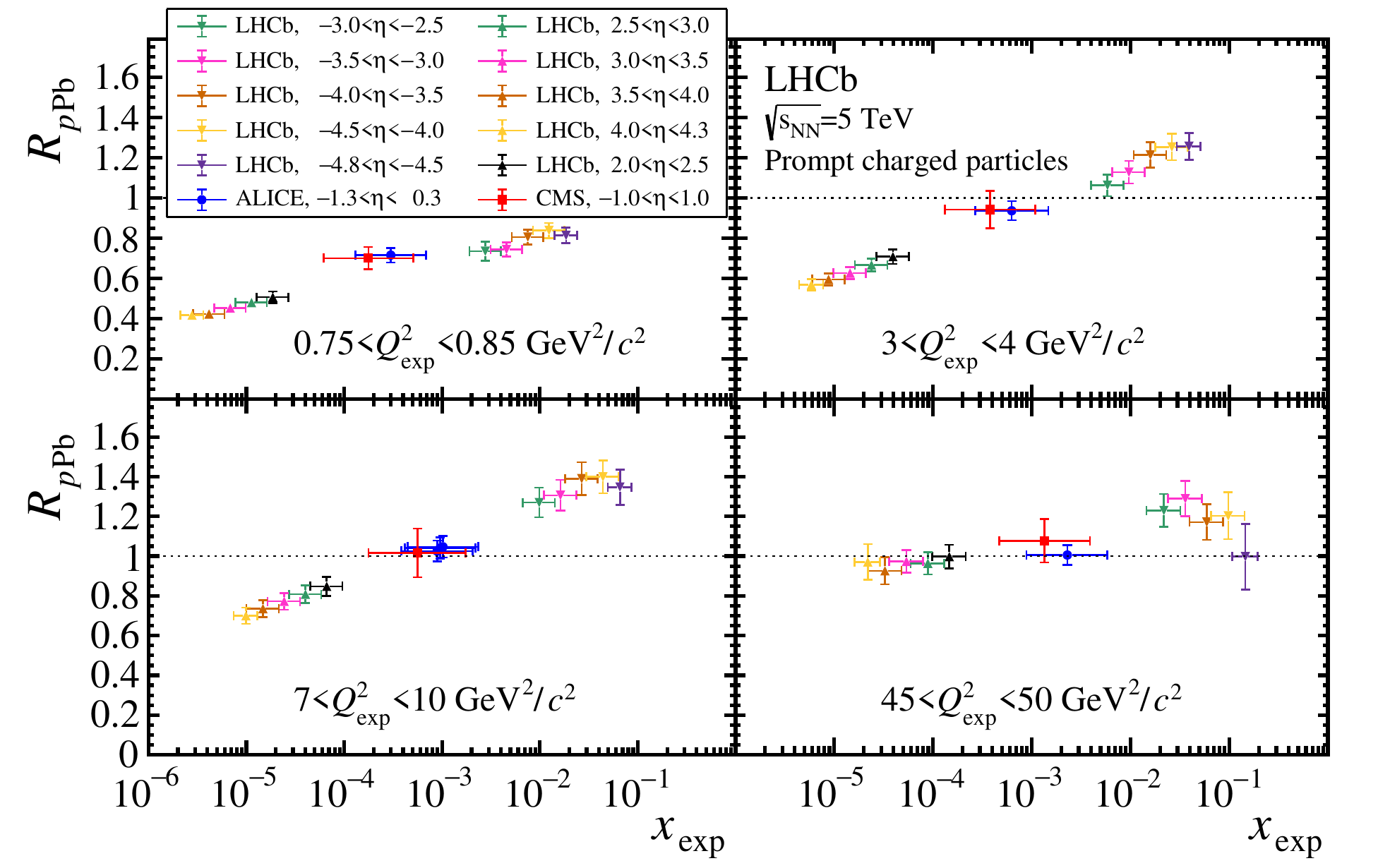}
\end{center}
\caption{Evolution of the nuclear modification factor with $x_{\text{exp}}$ from this study, ALICE~\cite{ALICEchpart2018}, and CMS~\cite{CMSchpart2017}, for  different $Q^{2}_{\text{exp}}$ ranges. Each plot includes all the $\Rppb$ ${(\eta,\pt)}$ intervals with a $\pt$ centre within the $Q^{2}_{\text{exp}}$ range specified in the plot. Horizontal error bars account for the minimum and maximum $x_{\text{exp}}$ value for a given ${(\eta,\pt)}$ interval. Vertical error bars correspond to statistical, systematic and luminosity (normalisation) uncertainties for LHCb (ALICE, CMS), added in quadrature.
}
\label{fig:RpAvsx_experiments}
\end{figure}

\Figrefb{fig:RpAvsx_experiments} shows the $\Rppb$ evolution with $x_{\text{exp}}$ for four $Q^{2}_{\text{exp}}$ intervals from this study and the results from the ALICE~\cite{ALICEchpart2018} and CMS~\cite{CMSchpart2017} collaborations. Since the $\pt$ binning is different among the three experiments, the $Q^{2}_{\text{exp}}$ ranges are selected to contain at least one $\pt$ interval from each experiment. A consistent trend between this measurement in the forward region, the measurements in the central region from ALICE and CMS and the result in the backward region is observed for the four $Q^{2}_{\text{exp}}$ intervals. The evolution of $\Rppb$ with $x_{\text{exp}}$ is $Q^2_{\text{exp}}$-dependent.

In summary, the differential production cross-sections have been measured in $\pt$ and $\eta$ intervals for prompt charged particles produced in \pp and \pA collisions at ${\sqsnn=5\tev}$. The measurement corresponds to ${\ptot>2\gevc}$ and ${0.2<\pt<8.0\gevc}$ prompt charged particles with $2.0<\eta<4.8$ in \pp and ${-5.2<\eta<-2.5}$ and ${1.5<\eta<4.3}$ in \pA collisions. This is the first determination of such cross-sections in \pA collisions in the forward and backward regions at the LHC, and the first measurement in \pp collisions at ${\sqsnn=5\tev}$. The total uncertainty is around $3\%$ for most kinematic intervals both in \pp and \pA collisions. As a result, the data place stringent constraints on non-perturbative QCD models in high-energy nuclear collisions.

The nuclear modification factor $\Rppb$ is also determined and is one of the most precise to date. The total uncertainty, including the normalisation contribution, is below $5\%$ for most of the ${(\eta,\pt)}$ intervals. In the forward region, a suppression of the charged particle production is observed, especially for low $\pt$ and the most forward $\eta$. In the backward region, the production of charged particles with $\pt>1.5\gevc$ is significantly enhanced. The $\Rppb$ shape exhibits a clear pseudorapidity dependence. These data cannot be simultaneously described across the entire measured $\eta$ range by nPDFs alone. Contrary to what is observed at central rapidity~\cite{ALICE:2021est}, the forward data are inconsistent with CGC calculations at the lowest $\pt$. Multiple scattering calculations, which successfully reproduce PHENIX results~\cite{PHENIXchpart2020}, fail to describe the backward region. These measurements provide strong constraints on nuclear PDFs at the lowest accessible $x$ ranges, and show that additional, previously unconsidered mechanisms are required to provide a consistent description of particle production in nuclear collisions at the LHC.
\clearpage

\section*{Acknowledgements}
%
We would like to thank to Hannu Paukkunen, Ilka Helenius, Heikki M\"antysaari and Zhong-Bo Kang, for providing LHCb specific theoretical predictions.
\noindent We express our gratitude to our colleagues in the CERN
accelerator departments for the excellent performance of the LHC. We
thank the technical and administrative staff at the LHCb
institutes.
We acknowledge support from CERN and from the national agencies:
CAPES, CNPq, FAPERJ and FINEP (Brazil); 
MOST and NSFC (China); 
CNRS/IN2P3 (France); 
BMBF, DFG and MPG (Germany); 
INFN (Italy); 
NWO (Netherlands); 
MNiSW and NCN (Poland); 
MEN/IFA (Romania); 
MSHE (Russia); 
MICINN (Spain); 
SNSF and SER (Switzerland); 
NASU (Ukraine); 
STFC (United Kingdom); 
DOE NP and NSF (USA).
We acknowledge the computing resources that are provided by CERN, IN2P3
(France), KIT and DESY (Germany), INFN (Italy), SURF (Netherlands),
PIC (Spain), GridPP (United Kingdom), RRCKI and Yandex
LLC (Russia), CSCS (Switzerland), IFIN-HH (Romania), CBPF (Brazil),
PL-GRID (Poland) and OSC (USA).
We are indebted to the communities behind the multiple open-source
software packages on which we depend.
Individual groups or members have received support from
AvH Foundation (Germany);
EPLANET, Marie Sk\l{}odowska-Curie Actions and ERC (European Union);
A*MIDEX, ANR, Labex P2IO and OCEVU, and R\'{e}gion Auvergne-Rh\^{o}ne-Alpes (France);
Key Research Program of Frontier Sciences of CAS, CAS PIFI,
Thousand Talents Program, and Sci. \& Tech. Program of Guangzhou (China);
RFBR, RSF and Yandex LLC (Russia);
GVA, XuntaGal and GENCAT (Spain);
the Royal Society
and the Leverhulme Trust (United Kingdom).

\addcontentsline{toc}{section}{References}
\bibliographystyle{LHCb}
\bibliography{main,standard,LHCb-PAPER,LHCb-CONF,LHCb-DP,LHCb-TDR}
 
\newpage
\centerline
{\large\bf LHCb collaboration}
\begin
{flushleft}
\small
R.~Aaij$^{32}$,
C.~Abell{\'a}n~Beteta$^{50}$,
T.~Ackernley$^{60}$,
B.~Adeva$^{46}$,
M.~Adinolfi$^{54}$,
H.~Afsharnia$^{9}$,
C.A.~Aidala$^{86}$,
S.~Aiola$^{25}$,
Z.~Ajaltouni$^{9}$,
S.~Akar$^{65}$
% ...

{\footnotesize \it

% ...
$^{9}$Universit{\'e} Clermont Auvergne, CNRS/IN2P3, LPC, Clermont-Ferrand, France\\
% ...
$^{25}$INFN Sezione di Milano, Milano, Italy\\
% ...
$^{32}$Nikhef National Institute for Subatomic Physics, Amsterdam, Netherlands\\
% ...
$^{46}$Instituto Galego de F{\'\i}sica de Altas Enerx{\'\i}as (IGFAE), Universidade de Santiago de Compostela, Santiago de Compostela, Spain\\
% ...
$^{50}$Physik-Institut, Universit{\"a}t Z{\"u}rich, Z{\"u}rich, Switzerland\\
% ...
$^{54}$H.H. Wills Physics Laboratory, University of Bristol, Bristol, United Kingdom\\
% ...
$^{60}$Oliver Lodge Laboratory, University of Liverpool, Liverpool, United Kingdom\\
% ...
$^{65}$University of Cincinnati, Cincinnati, OH, United States\\
% ...
$^{68}$Syracuse University, Syracuse, NY, United States\\
% ...
$^{86}$University of Michigan, Ann Arbor, United States, associated to $^{68}$\\
\bigskip
% ...
\medskip
}
\end{flushleft}

\end{document}